\documentstyle[11pt]{article}

\newcommand{\K}{\mbox{ , }}
\newcommand{\p}{\mbox{ .}}

\newcommand{\ppT}{\frac{\partial p}{\partial T}}
 
\renewcommand{\epsilon}{\varepsilon}

\begin{document}
\title{Bulk Viscous Cosmology}
\author{Winfried Zimdahl\thanks{Present address: Fakult\"{a}t  
f\"{u}r Physik, Universit\"{a}t Konstanz, PF 5560 M678, D-78434  
Konstanz, Germany}\\ 
Departament de F\'{\i}sica,
Universitat Aut\`{o}noma de Barcelona\\
E-08193 Bellaterra (Barcelona), Spain\\}
\date{\today}
\maketitle
\begin{abstract}
The full causal M\"uller-Israel-Stewart (MIS) theory 
of dissipative processes in relativistic fluids is
applied to a flat, homogeneous and isotropic universe with bulk  
viscosity. 
It is clarified in which sense the so called truncated version is a
reasonable limiting case of the full theory.  
The possibility of bulk viscosity driven inflationary solutions of
the full theory is discussed. 
As long as the particle number is conserved almost all 
these solutions exhibit an exponential increase of the temperature. 
Assuming that the bulk viscous pressure of the MIS theory
may also be interpreted as an effective description for particle  
production
processes, the thermodynamical behaviour of the Universe changes
considerably. 
In the latter case the temperature increases at a lower rate or 
may remain constant during a
hypothetical de Sitter stage, accompanied by a substantial growth of
the comoving entropy.
\end{abstract}
\newpage
\section{Introduction}
Nonequilibrium thermodynamical processes are supposed to play a
crucial role in the physics of the early Universe. 
Traditionally, for the description of these phenomena the theories of
Eckart [1] and Landau and Lifshitz [2] were used. 
Due to the work of M\"uller [3], Israel [4], Israel and Stewart [5,
6], Pav\'on, Jou and Casas-V\'azquez [7], 
Hiscock and Lindblom [8] it became clear however, that the Eckart
type theories suffer from serious drawbacks concerning causality and
stability. These difficulties could be traced back to their
restriction to first order deviations from equilibrium. 
If one includes higher order deviations as well, the corresponding
problems disappear. 
By now it is generally agreed that any analysis of dissipative
phenomena in relativity should be based on the theories
of M\"uller, Israel and Stewart, including at least second order
deviations from equilibrium, 
although in specific cases the
latter might reproduce results of the Eckart theory [9]. 
Cosmological implications of second order theories were first
considered by Belinskii et al. [10]. 
In the realm of cosmology especially bulk viscous phenomena have
attracted considerable interest (see, e.g., [11]) since
bulk viscosity is the only possible dissipative mechanism in
homogeneous and isotropic spacetimes. 
While the coefficient of bulk viscosity vanishes both for  pure
relativistic and  pure nonrelativistic 
equations of state, it may be important,
e.g., for mixtures of radiation and matter [12]. 
On the other hand, it is well known [13-15] and widely used [17-25] that
particle production processes in the expanding Universe may be
phenomenologically described  in terms of effective viscous
pressures. 

A major point of interest in the study of bulk viscous universes has
been the question whether there are conditions under
which a sufficiently large bulk viscous pressures could
lead to an inflationary behaviour. 
While some authors concluded that a bulk viscosity driven inflation
is impossible [26], others [27-29] found  inflationary
solutions of the cosmological evolution equations. 
Partially, these differences occur since different equations of state
were used. 
But the results also depend on whether the investigations were
performed within the full, causal second order theory or in a
truncated version of the latter. 

One should be aware that discussing the issue of bulk viscous
driven inflationary solutions at all, implies in any case an
extrapolation of nonequilibrium thermodynamical theories beyond the
range for which their applicability was strictly justified [29]. 
Bulk viscous inflation, if it exists, is a far-from-equilibrium 
phenomenon, while even the 
full causal second order 
MIS theory is a theory for small deviations from
equilibrium. Therefore all theoretical conclusions are necessarily
tentative. It is the hope that they nevertheless will provide an
indication of the correct behaviour far from equilibrium. 

As to the relation between the full theory and the truncated version,
to be discussed in some detail below, the following comment should be
made from the outset. 
To decide whether a theory is truncated or not on apparently obvious
formal grounds, i.e., from the appearance of the causal evolution
equations may be misleading. 
The structure of the evolution equation depends on the choice of the
basic thermodynamical variables. 
In most cases the latter are equilibrium variables and the Gibbs
equation has their familiar form. In the framework of `Extended
Irreversible Thermodynamics' (EIT) however, a generalized Gibbs  
equation is
used which includes dissipative quantities as independent variables. 
Temperature and pressure in EIT are nonequilibrium quantities,
different from their equilibrium counterparts. 
Written in terms of these nonequilibrium quantities the causal
evolution equation formally may look identical to the truncated
theory written in terms of the more familiar equilibrium variables. 
These points have been clarified in a recent paper by Gariel and Le
Denmat [30] who pointed out that the apparently (because of the use
of nonequilibrium variables) truncated theory of Pav\'on et al. [7]
in fact is equivalent to the full theory of Israel [4]. 
The reader is also referred to corresponding comments in [28] and
[29]. 

In the present paper the symbols $T$ and $p$ always denote the
equilibrium temperature and the equilibrium pressure, respectively. 

In section 2 we reconsider the bulk viscous cosmological dynamics
within the  causal second order theories. 
The evolution law for the temperature of the cosmic fluid is
shown to be different in general 
from the ad hoc relationship used in previous
treatments by Romano and Pav\'{o}n [31], Zakari and Jou [28] and  
Maartens
[29] (subsection 2.1). 
In subsection 2.2 we discuss the conditions under which the truncated
version yields results close or identical to those of the full
theory. 
The general viscous fluid dynamics of the full MIS theory is
presented in subsection 2.3. 
Subsection 2.4 investigates the conditions for viscous exponential
inflation. Almost all corresponding solutions  
imply an exponential increase
of the fluid temperature.  
As a consequence of this behaviour there is in general no
substantial growth of the comoving entropy as was found previously  
[29]. 
The latter result corresponds to a very specific limiting case. 
In section 3 the viscous pressure of the full causal theory is
assumed to describe partially or fully the effect of particle
creation taken into account by 
a nonvanishing source term in the particle number
balance. 
In this setting the backreaction of the viscous pressure on the
temperature is different from the conventional viscous fluid case of
section 2. In the limit that the viscous pressure is entirely due to
particle production there exist stable inflationary solutions for
which both the particle number density and the temperature are
constant and, moreover, the comoving entropy grows exponentially. 
Section 4 summarizes the results of the paper. 
Units have been chosen so that $c = k_{B} = 1$. 
\section{Bulk viscous fluid dynamics}
\subsection{General relations}
The energy momentum tensor of a relativistic fluid with bulk
viscosity as the only dissipative phenomenon is 
\begin{equation}
T^{ik} = \rho u^{i}u^{k} 
+ \left(p + \pi\right) h^{ik}\p \label{1}
\end{equation}
$\rho$ is the energy density,   
$u^{i}$ is the 4-velocity, $p$ is the equilibrium pressure, 
$h^{ik}$ is the
projection tensor 
$h^{ik} =g^{ik} + u^{i}u^{k}$, and 
$\pi$ is the bulk viscous pressure.\\
The particle flow vector $N^{a}$ is given by 
\begin{equation}
N^{a} = nu^{a} \K\label{2}
\end{equation}
where  $n$ is the
particle number density. 
Limiting ourselves to second order deviations from equilibrium, the
entropy flow vector $S^{a}$ takes the form [5, 26]
\begin{equation}
S^{a} = sN^{a} 
- \frac{\tau\pi^2}{2\zeta T} u^{a}\p\label{3}
\end{equation}
$s$ is the entropy per particle, $\tau$ is the relaxation time, 
$T$ is the temperature and $\zeta$ is the coefficient of bulk
viscosity. 
The conservation laws 
\begin{equation}
N^{a}_{;a}  =  0\K\label{4}
\end{equation}
and 
\begin{equation}
T^{ab}_{\ ;b}  =  0\K\label{5}
\end{equation}
imply 
\begin{equation}
\dot{n} + \Theta n =  0\K\label{6}
\end{equation}
and 
\begin{equation}
\dot{\rho} = - \Theta\left(\rho + p + \pi\right)\K
\label{7}
\end{equation}
respectively, where $\Theta \equiv u^{a}_{;a}$ is the fluid  
expansion and 
$\dot{n} \equiv n_{,a}u^{a}$ etc. 
Combining (\ref{6}) and (\ref{7}) 
with the Gibbs relation 
\begin{equation}
Tds = d\frac{\rho}{n} + pd\frac{1}{n} \K\label{8}
\end{equation}
we get 
\begin{equation}
nT\dot{s} =  - \Theta\pi
\p\label{9}
\end{equation}
From (\ref{3}) and (\ref{6})  we find
\begin{equation}
S^{a}_{;a}  =  - \frac{\pi}{T}\left[\Theta + \frac{\tau}{\zeta}
\dot{\pi} + \frac{1}{2}\pi T\left(\frac{\tau}{\zeta T}u^{a}\right)_{;a}
\right]
\label{10}
\end{equation}
for the entropy production density $S^{a}_{;a}$. 
The simplest way to guarantee  $S^{a}_{;a} \geq 0$ implies the evolution
equation
\begin{equation}
\pi + \tau\dot{\pi}  =  - \zeta \Theta - 
\frac{1}{2}\pi \tau\left[\Theta + \frac{\dot{\tau}}{\tau} 
- \frac{\dot{\zeta}}{\zeta} - \frac{\dot{T}}{T} 
\right] \label{11}
\end{equation}
for $\pi$, leading to 
\begin{equation}
S^{a}_{;a}  =  \frac{\pi^{2}}{\zeta T}
\p\label{12}
\end{equation}
For $\tau \rightarrow 0$ eq.(\ref{11}) reduces to the corresponding  
relation of
the Eckart theory. 
The frequently used truncated version 
\begin{equation}
\pi + \tau\dot{\pi}  =  - \zeta \Theta 
\K\label{13}
\end{equation}
also known as Maxwell-Cattaneo equation, 
follows  if the bracket term on the r.h.s of  (\ref{11}) can be
neglected compared with the viscosity term $- \zeta \Theta$. Below we
shall give explicit criteria for this approximation. 
If the bracket term vanishes identically, i.e., if  the condition  
\begin{equation}
\Theta + \frac{\dot{\tau}}{\tau} 
- \frac{\dot{\zeta}}{\zeta} - \frac{\dot{T}}{T} = 0
\label{14}
\end{equation} 
is fulfilled, the full and the truncated theories become identical. 
As we shall see this is possible only in exceptional cases. 

Let us assume equations of state in the general form
\begin{equation}
p = p\left(n, T\right) \label{15}
\end{equation}
and
\begin{equation}
\rho = \rho \left(n,T\right)\K \label{16}
\end{equation}
according to which the particle number density $n$ and the
temperature $T$ are our basic thermodynamical variables. 
Differentiating the latter relation, 
using the balances (\ref{6}) and (\ref{7}) as well as the general  
thermodynamic
relation 
\begin{equation}
\frac{\partial \rho}{\partial n} = 
\frac{\rho + p}{n} 
- \frac{T}{n}\ppT \K \label{17}
\end{equation}
one finds the following evolution law for the temperature (cf. [22]): 
\begin{equation}
\dot{T}  =  \Theta n \frac{\partial \rho/\partial n}
{\partial \rho/\partial T} 
+  \frac{\dot{\rho}}{\partial \rho/\partial T} 
\K  \label{17a}
\end{equation}
or
\begin{equation}
\frac{\dot{T}}{T}  = - \Theta
\left[\frac{\partial p/\partial T}{\partial \rho/\partial T} 
+  \frac{\pi}{T \partial \rho/\partial T}\right]
\p  \label{18}
\end{equation}
For $\pi = 0$ and with $\Theta = 3\dot{R}/R$, where $R$ is the scale
factor of the Robertson-Walker metric, 
(\ref{18}) reproduces 
the well known $T_{r} \sim R^{-1}$ behaviour in a radiation 
dominated Friedmann-Lema\^{\i}tre-Robertson-Walker (FLRW) Universe, 
while for $\rho = nm + \frac{3}{2}nT$,  
$p = nT$  and $T \ll m$ we recover  
$T_{m} \sim R^{-2}$ in the matter dominated case. 
For a viscous fluid  the behaviour of the
temperature depends on $\pi$. 
Since $\pi$ is expected to be negative, the second term in the
bracket on the r.h.s. of (\ref{18}) will counteract the first one. 
Close to equilibrium, i.e., for $\mid \pi\mid < p$ the existence of a
bulk viscous pressure implies that in an expanding universe 
the temperature decreases 
less rapidly than in the perfect fluid case. 
\subsection{The truncated version}
While the truncated version was used in most of the earlier
applications, more recently an increasing number of authors [26, 28,
31, 29] has studied the full theory and compared the results of the
latter wih those of the truncated version. 
In some cases these results differ dramatically, which may be
interpreted as a breakdown of the Maxwell-Cattaneo type equations as
a reasonable approximation to the full theory under the corresponding
conditions. What seems to be missing, however, are general criteria
according to which one may decide whether the truncated version is
sensible and beyond which limits it fails to give an answer close to
that of the full theory. 
Intuitively one expects the coincidence to be the better the closer
one is to the equilibrium case. 
Below we shall give an example showing that there are identical
results in exceptional cases even far from equilibrium. 
In order to clarify the approximative character of the truncated
theory we assume, as usual, the relation $\zeta = \rho \tau$  that
guarantees a finite propagation velocity of viscous pulses [10,
27-29, 31]. In the following subsection we are going to generalize
this relation. 
Using  (\ref{7}) and (\ref{18}) 
in equation (\ref{11}) with $\zeta = \rho \tau$ 
we find 
\begin{equation}
\pi + \tau\dot{\pi}  =  - \rho\Theta\tau\left[1 +  
\frac{\pi}{2\rho}\left(1 + b + \gamma\right)   
+ \frac{\pi^{2}}{2\rho^{2}} \left(a + 1\right)\right]
\K\label{19}
\end{equation}
with the abbreviations 
\begin{equation}
a \equiv  \frac{\rho}{nTc_{v}} \K 
c_{v} \equiv  \frac{1}{n}\frac{\partial \rho}{\partial T}  \K 
b \equiv  \frac{\partial p/\partial T}{\partial \rho/\partial T} = 
\frac{1}{nc_{v}}\frac{\partial p}{\partial T}  
\K \gamma = \frac{\rho + p}{\rho} 
\p\label{20}
\end{equation}
Obviously, the truncated version is expected to be applicable for
\begin{equation}
 \frac{\mid\pi\mid}{2\rho} \left(1 + b + \gamma\right)  \ll 1 \K 
\frac{\pi^{2}}{2\rho^{2}}\left(1 + a\right) \ll 1  
\p\label{21}
\end{equation}
Since for `ordinary' matter $b$ and $\gamma$ lie in the ranges 
$1/3 \leq b \leq 2/3$ and $1 \leq \gamma \leq 4/3$, respectively, the
first condition is roughly equivalent to $\pi \ll \rho$. 
For radiation with $\gamma = 4/3$, $b = 1/3$ and $a = 1$ the second
condition is implied by the first one. 
For matter with $\gamma = 1$, $b = 2/3$ and $a = \frac{2m}{3T} \gg 1$
however, the second condition has to be checked separately. 
The inequalities (\ref{21}) may be regarded as criteria under which 
the truncated theory is a reasonable approximation to the full
theory. 
Given equations of state (\ref{15}) and (\ref{16}), any solution
$\pi$ of the truncated theory may be tested according to (\ref{21}) 
whether or not and to which accuracy it approximates the full theory.

But there is a different possibility, namely the case 
\begin{equation}
\frac{\pi}{2\rho}\left(1 + b + \gamma\right)   
+ \frac{\pi^{2}}{2\rho^{2}} \left(a + 1\right) = 0
\K\label{22}
\end{equation} 
in which all the terms that distinguish the full from the truncated
theory cancel among themselves. 
Relation (\ref{22}) is identical to (\ref{14}) for $\zeta =  
\rho\tau$, i.e., 
it is the condition under which the full theory is identical to the
truncated one. 
Solving (\ref{22}) for $\pi/\rho$ yields 
\begin{equation}
\frac{\pi}{\rho} = -  \frac{1 + \gamma + b}{1 + a} 
\p\label{23}
\end{equation}
According to the above mentioned parameter ranges for $\gamma$, $b$
and $a$, $\mid \pi\mid \ll \rho$ is only possible for $a \gg 1$,
i.e., for massive particles. 
Using (\ref{23}) in (\ref{7}) yields 
\begin{equation}
\frac{\dot{\rho}}{\rho}  = - \Theta \frac{\gamma a - 1 - b}{1 + a}
\p  \label{24}
\end{equation}

For radiation ($\gamma = 4/3$, $b = 1/3$, $a = 1$) we find 
$\dot{\rho} = 0$ and, consequently, $\Theta = \Theta_{0} = const$ in
a flat FLRW universe. 
The truncated and the full theory coincide in a specific bulk
viscosity driven inflationary universe. 
Since with (\ref{23}) $\pi$ is completely determined by $\rho$,
provided, the equations of state are given, the remaining equation 
(\ref{13}) is no longer a dynamical equation on its own, 
but may be used to calculate $\tau = \tau\left(\Theta_{0}\right)$. 
Since the solution is stationary, i.e., $\dot{\pi} = 0$, we find 
\begin{equation}
\tau_{0}^{r}  =  \frac{4}{3}\Theta_{0}^{-1}
\p  \label{25}
\end{equation}
In a radiation dominated universe the full theory and the truncated
version admit a common, bulk viscosity driven inflationary solution
with a relaxation time of the order of the expansion time. 
This seems to be a new result. 
For the temperature dependence we find from (\ref{18}) and (\ref{23}) 
\begin{equation}
\frac{\dot{T}}{T}  = - \Theta \frac{b - a\left(\gamma + 1\right)}{1 + a}
\K  \label{26}
\end{equation}
yielding 
\begin{equation}
\frac{\dot{T}}{T}  =  \Theta 
  \label{27}
\end{equation}
for radiation, or $T \sim R^{3}$. 
The temperature increases in an expanding universe. This implication
of the condition (\ref{14}) was first noticed by Maartens [29]. 
While this might appear strange at the first glance, it is an
unavoidable feature of any bulk viscosity driven inflation as is
obvious from (\ref{18}). 
This point will be discussed in more detail in the following
subsection. 
\subsection{The dynamics of the full second order theory}
In this subsection we shall investigate the full causal theory  
assuming the
existence of general equations of state (\ref{15}), (\ref{16}) and 
\begin{equation}
\frac{\zeta}{\tau} \equiv f  =  f\left(\rho\right)  
\p  \label{28}
\end{equation}
Following Belinskii et al.[10] usually 
the relation $\zeta/\tau = \rho$ was used to guarantee that the
propagation velocity of viscous pulses, which is expected to be of the
order [28] 
\begin{equation}
v \sim \left(\frac{\zeta}{\rho\tau}\right)^{1/2} \K \label{29}
\end{equation}
does not exceed the velocity of light. 
We shall not immediately specify to $f = \rho$ 
in order to admit a certain range for this propagation velocity. 
As we shall see below, this additional freedom, allowing, e.g., 
$f = \alpha \rho$ with $0 < \alpha \leq 1$, may be useful in  
characterizing 
a possible inflationary phase. 
In this case one has $\dot{f} = f'\dot{\rho}$, where 
$f' \equiv df/d\rho$. 
Consequently, using (\ref{7})
\begin{equation}
\frac{\dot{f}}{f}  = - \Theta\rho \frac{f'}{f}
\left(\gamma + \frac{\pi}{\rho}\right)
\p  \label{30}
\end{equation}
Together with (\ref{18}), 
equation (\ref{11}) may now be written as
\begin{equation}
\pi + \tau\dot{\pi}  =  - \rho\Theta\tau\left[\frac{f}{\rho} +  
\frac{\pi}{2\rho}\left(1 + b + \gamma\frac{\rho f'}{f}\right)   
+ \frac{\pi^{2}}{2\rho^{2}} \left(a + \frac{\rho f'}{f}\right)\right]
\p\label{31}
\end{equation}
It is obvious how the applicability conditions (\ref{21}) of the
truncated version have to be modified in this more general case. 
Restricting ourselves to a flat FLRW universe with  
\begin{equation}
\frac{\Theta^{2}}{3}  = \kappa\rho 
\K  \label{32}
\end{equation}
where $\kappa$ is Einstein's gravitational constant, 
and 
\begin{equation}
\dot{\Theta}  = - \frac{3\kappa}{2}\left(\rho + p + \pi\right)
\K  \label{33}
\end{equation}
the latter equation may be used to eliminate $\pi$ and $\dot{\pi}$,
respectively. 
The bulk pressure $\pi$ may be written as
\begin{equation}
\kappa\pi = - 3\gamma H^{2} - 2\dot{H} \K\label{34}
\end{equation}
where $3 H = \Theta$ and $H \equiv \dot{R}/R$ or, 
\begin{equation}
\frac{\pi}{\rho} = - \gamma -  
\frac{2}{3}\frac{\dot{H}}{H^{2}}\p\label{35}
\end{equation}
From (\ref{15}), using (\ref{6}) and (\ref{18}) one realizes
\begin{equation}
\dot{p} = v_{s}^{2}\dot{\rho} + \Theta \pi\left(v_{s}^{2} - 
b\right)\K \label{36} 
\end{equation}
with the sound velocity $v_{s}$, given by
\begin{equation}
v_{s}^{2} = 
\left(\frac{\partial p}{\partial \rho}\right)_{isentropic}
= \frac{n}{\rho + p}
\frac{\partial p}{\partial
n} + \frac{T}{\rho + p} 
\frac{\left(\partial p/\partial T\right)^{2}}
{\partial \rho/\partial T}
 \p \label{37} 
\end{equation}
Differentiating 
(\ref{33}) and using (\ref{7}) and (\ref{36}), 
$\kappa\dot{\pi}$ on the l.h.s. of (\ref{31}) may be replaced by
\begin{equation}
\kappa\dot{\pi} = - 2 \ddot{H} - 6 H\dot{H}\left(1 + b\right) 
+ 9H^{3}\gamma 
\left(v_{s}^{2} - b\right)\p \label{38} 
\end{equation}
Applying (\ref{34}) and (\ref{38}) in (\ref{31}) we arrive at the  
follwing
evolution equation for $H$: 
\begin{eqnarray}
\tau \ddot{H} - \frac{\dot{H}^{2}}{H}\tau\left(a +
\frac{\rho}{f}f'\right) 
- 3\dot{H}H\tau\left[\gamma a 
+ \frac{\gamma}{2} \frac{\rho}{f}f' - \frac{3}{2}\left(1 +  
b\right)\right]&&
\nonumber\\
+ \dot{H} - \frac{9}{2}H^{3}\tau\left[\frac{f}{\rho} 
+ \frac{\gamma}{2}\left(\gamma a - 1 - b\right)  
+ \gamma\left(v_{s}^{2} - b\right)\right] 
+ \frac{3}{2}\gamma H^{2}&=& 0\p \label{39}
\end{eqnarray}
It should be pointed out again that in arriving at this evolution
equation  only equations of state (\ref{15}), (\ref{16}) and
(\ref{28})  were used. 
We think this set of equations of state to be more general than 
those used in previous papers [31, 28, 29]. 
Especially, no specific temperature law like $T = \beta\rho^{r}$ 
in [31] had to be postulated. 
There is no freedom to impose a separate temperature law. 
The behaviour of the temperature is generally governed by 
(\ref{18}). 
As it is obvious from (\ref{17a}), a relation $T \sim \rho^{r}$ 
is possible for $\partial \rho/\partial n = 0$. 
Even in this case, however, $r$ is not arbitrary but determined by 
$r = \rho \left(T d \rho/ dT\right)^{-1}$. 
In the specific case of a flat FRLW universe with 
\begin{equation}
\frac{\dot{H}}{H} = \frac{1}{2}\frac{\dot{\rho}}{\rho} \label{40}
\end{equation}
equation (\ref{18}) reduces to 
\begin{equation}
\frac{\dot{T}}{T} = 3\left(\gamma a - b\right)\frac{\dot{R}}{R} 
+ a \frac{\dot{\rho}}{\rho} \p\label{41}
\end{equation}
For constant values of $a$, $b$ and $\gamma$ 
\begin{equation}
T \sim \rho^{a}R^{3\left(\gamma a - b\right)} \label{42}
\end{equation}
results, which in the radiation dominated case 
with $a = 1$, $b = 1/3$, 
$\gamma = 4/3$, specifies to 
\begin{equation}
T_{r} \sim \rho_{r}R^{3}\p \label{43}
\end{equation}
Obviously, with $\rho_{r} \sim R^{-4}$ for $\pi = 0$, one obtains 
the correct limiting case $T_{r} \sim R^{-1}$. 
\subsection{Viscous exponential inflation}
It has been a matter of some debate whether a bulk viscous pressure is
able to drive inflation [26-29, 31]. 
While Hiscock and Salmonson [26] have shown that, using the equations of
state for a Boltzmann gas, the full causal theory does not admit an
inflationary phase, different equations of state like those used by 
Romano and Pav\'on [31], Zakari and Jou [28] and Maartens [29] 
are compatible with a de
Sitter phase. 
It should be stressed again that 
dealing with
the question of a viscosity driven inflation one has to assume that
the MIS theory is applicable far from equilibrium [29]. 
Under these premises we are going to consider now the possibility of a
solution of the evolution equation for $H$ with $H = H_{0} = const$. 
In such a case eq.(\ref{39}) yields 
\begin{equation}
\tau_{0}^{-1} = 3 H_{0}
\left[\frac{f}{\gamma\rho} + \frac{1}{2}\left(\gamma a - 1 - b\right) 
+ v_{s}^{2} - b\right]\p\label{44}
\end{equation}
In the radiation dominated case this reduces to
\begin{equation}
\tau_{0}^{r} = \frac{4}{9}\frac{\rho}{f}H^{-1}_{0}\p\label{45}
\end{equation}
Keeping in mind that until now only $f = f\left(\rho\right)$ was  
used, one
recognizes that the ratio between $\tau^{r}_{0}$ and $H_{0}^{-1}$  
crucially
depends on $f/\rho$. 
If, as is usually assumed, $f = \rho$, we have $\tau^{r}_{0} <  
H_{0}^{-1}$. 
If, however, $f$ is allowed to be smaller than $\rho$, we may have 
$\tau^{r}_{0} > H_{0}^{-1}$. It follows from (\ref{29}) that a relation 
$f < \rho$ implies a lower propagation velocity compared with the
case $f = \rho $. Provided $f = \rho $ is equivalent to a propagation
velocity that coincides with the velocity of light, $f = \rho/3$, 
e.g., leads to a propagation with $v_{s} = 1/\sqrt{3}$, the velocity
of sound (see Israel and Stewart [6]).  
In the latter case the relation (\ref{45}) specifies to 
\begin{equation}
\tau_{0}^{r} = \frac{4}{3}H_{0}^{-1}\K\label{46}
\end{equation}
implying $\tau_{0}^{r} > H_{0}^{-1}$. 
In this case the nonequilibrium is `frozen in'. The viscous inflation is
`nonthermalizing' [29]. 
While for $\tau^{r}_{0} < H_{0}^{-1}$ there is a quick (compared  
with the
expansion rate) relaxation to (local) equilibrium, this is no longer the
case if $\tau^{r}_{0}$ becomes comparable to the expansion rate  
$H_{0}^{-1}$. 
Nonequilibrium situations like this may occur, e.g., in GUTs close  
to the
Planck time where the underlying microscopic process is the decay of
heavy vector bosons [32]. 
According to Maartens [29] the condition $\tau_{0} > H_{0}^{-1}$ may be
regarded as a consistency criterion for causal viscous inflation. 
Clearly, for $\tau_{0} < H_{0}^{-1}$, the Universe will relax to an
equilibrium state in less than one expansion time. 
A successful inflation, however, 
has to last for many characteristic
expansion times. 
 
In the matter dominated case with $\gamma = 1$, $b = 2/3$, 
$v_{s}^{2} \ll 1$, $a = \frac{2m}{3T} \gg 1$, we find 
\begin{equation}
\tau_{0}^{m} \sim \frac{T}{m}H_{0}^{-1}\K\label{47}
\end{equation}
i.e., we are always in the range $\tau^{m}_{0} \ll H_{0}^{-1}$. 
The time scale for the relaxation to equilibrium is much smaller than
the expansion time scale. 

Calculating the dependence $\tau_{0}\left(H_{0}\right)$ along the
the same lines that lead to (\ref{44})
within the truncated version provides us
with 
\begin{equation}
\tau_{0}^{-1} = 3\gamma^{-1}
\left[\frac{f}{\rho}  
+ \gamma\left(v_{s}^{2} - b\right)\right]H_{0}\p\label{48}
\end{equation}
Specializing to the radiation dominated case, (\ref{48}) reduces to
(\ref{45}), i.e., the truncated version yields exactly the same  
dependence
as the full theory. 
We have rediscovered the specific case, discussed in subsection 2.2,
in which the full theory coincides with the truncated version.
For the matter dominated case, however, the result of the truncated
theory is 
different from (\ref{47}). 
This is not unexpected as well after the considerations of subsection
2.2. 

We shall now look at the behaviour of the
basic thermodynamic quantities under the condition 
$H = H_{0}$. 
Because of (\ref{39}) this implies $\rho = \rho_{0} =
const$ and (\ref{35}) reduces to 
\begin{equation}
\frac{\pi}{\rho} = - \gamma \p \label{49}
\end{equation}
According to (\ref{6}) the particle number density decreases
exponentially. The temperature evolution law (\ref{18}) together with
(\ref{49}) becomes
\begin{equation}
\frac{\dot{T}}{T}  = -  3H_{0}
\left[\frac{\partial p/\partial T}{\partial \rho/\partial T} 
-  \frac{\rho}{nTc_{v}}\gamma\right]
\K  \label{50}
\end{equation}
equivalent to
\begin{equation}
\dot{T}  = \frac{3H_{0}}{c_{v}}\frac{\partial \rho}{\partial n} 
\p  \label{52}
\end{equation}
Generally, the temperature is not constant in the inflationary phase. 
Provided $c_{v}$ is finite,  
a constant temperature 
$T = T_{0}$ is only possible for 
$\partial \rho/\partial n = 0$, corresponding to 
$T \sim \rho^{r}$ with  
$r = \rho \left(T d \rho/ dT\right)^{-1}$ (see the discussion below 
(\ref{39})). 
For any $\partial \rho/\partial n > 0$ the temperature  $T$
increases during inflation. 
Since $\partial \rho/\partial n > 0$ and $\partial \rho/\partial n =
0$, respectively, lead to dramatically different results for the
entropy production during a bulk viscosity driven inflationary phase
(see below), this point deserves a detailed discussion. 
It is frequently used that in the equations of state for radiation in
equilibrium, $p = nT$ and $\rho = 3nT$, which are specific cases of 
(\ref{15}) and (\ref{16}), the number density $n$ may be eliminated
according to $n \sim T^{3}$, resulting in $p = p\left(T\right)$ and 
$\rho = \rho\left(T\right)$ with $p = \rho/3$ and $\rho \sim T^{4}$. 
Then $T$ is the only independent variable and 
$\partial\rho/\partial T \rightarrow d\rho/d T$. 
In order to check whether a corresponding procedure is possible for 
$\pi \neq 0$, we shall assume an arbitrary dependence 
$n = n\left(T\right)$ instead of the equilibrium relation 
$n \sim T^{3}$. 
With (\ref{18}) we find 
\begin{equation}
\dot{n}  = - 3 H T \frac{d n}{d T}
\left(b + a\frac{\pi}{\rho}\right)  
\p  \label{52a}
\end{equation}
Comparison with (\ref{6}) leads to the general relation
\begin{equation}
\frac{d T}{T} = \left(b + a\frac{\pi}{\rho}\right) \frac{d n}{n}
\p  \label{52b}
\end{equation}
For $\pi = 0$ we recover $T \sim n^{1/3}$ for radiation ($b = 1/3$). 
For matter ($b = 2/3$) the correct result $T \sim n^{2/3} \sim
R^{-2}$ is obtained as well.

With the condition (\ref{49}) for bulk viscosity driven exponential
inflation and for $a = const$  equation (\ref{52b}) yields 
\begin{equation}
T \sim n^{b - \gamma a}
\p  \label{52c}
\end{equation}
In the radiation dominated case 
\begin{equation}
n \sim T^{- 1}
  \label{52d}
\end{equation}
results.  
While in equilibrium $n$ is an increasing function of $T$, $n$
decreases with $T$ in a de Sitter phase, characterized 
by (\ref{49}), which is a far-from-equilibrium state. 
According to (\ref{52d}) the exponential decrease of the particle
number density following from (\ref{6}) neccessarily is accompanied
by a corresponding increase in the temperature. 
This is obviously incompatible with $T \sim \rho^{1/4}$ from  
$\partial \rho/\partial n = 0$. 
Equations of state with $\partial \rho/\partial n = 0$ 
imply that the exponential dilution of the
particles of the out-of-equilibrium fluid according to (\ref{6}) does
not have any impact at all on the energy density of this fluid. 
We conclude that the previously  [31, 28, 29] used 
ad hoc 
assumption $T \sim \rho^{r}$ 
is not consistent with (\ref{15}), (\ref{16}) and $n =
n\left(T\right)$ if applied to a bulk pressure driven inflationary
phase. 

While an increasing temperature during the de Sitter stage  appears 
unfamiliar, this kind of behaviour is not quite unexpected for an
equation of state (\ref{16}). Since $n$ decreases
exponentially, $T$ must increase accordingly in order to guarantee 
$\dot{\rho} = 0$, as long as $\partial \rho/\partial n > 0$ and 
$\partial \rho/\partial T > 0$. 
Exactly this kind of behaviour was found by Hiscock and Salmonson  
[26] for
the truncated Israel-Stewart theory. 
As was demonstrated above, the full theory exhibits a corresponding
feature as well. (Note, that according to [26] there does not exist a de
Sitter phase in the full theory for the case of a Boltzmann gas). 
If one does not use, however, a relationship $T \sim \rho^{r}$ 
and the 
behaviour of $T$ is governed by (\ref{52}) 
with $\partial \rho/\partial n > 0$, this has important
consequences for the entropy production 
during bulk viscous driven inflation, discussed by Maartens [29]. 
The entropy in a comoving volume is $\Sigma = nsR^{3}$. 
With (\ref{6}) and (\ref{9}) the change of $\Sigma$ is 
\begin{equation}
\dot{\Sigma} = - \frac{\pi\Theta}{T}R^{3}\p\label{53}
\end{equation}
In the de Sitter phase with (\ref{49}) the latter expression reduces to 
\begin{equation}
\dot{\Sigma}_{H = H_{0}} = 3H_{0} \gamma 
\frac{\rho_{0}R^{3}}{T}\p\label{54}
\end{equation}
This change of the comoving entropy depends crucially on the
behaviour of $T$. 
For $T = const$ which follows from $T \sim \rho^{r}$, equivalent to 
$\partial \rho/\partial n = 0$ (see (\ref{52})), we recover the
exponential increase of $\dot{\Sigma}$ found by Maartens [29]. 
For equations of state (\ref{15}) and  (\ref{16}) however, with 
$\partial \rho/\partial n > 0$, the temperature dependence in the
case of radiation ($\gamma = 4/3$) is $T_{r} \sim \rho_{0} R^{3}$ as
follows from (\ref{43}).  
$\dot{\Sigma}$ turns out to be constant rather than 
exponentially increasing. 
Integrating (\ref{54}) one finds
only a linear growth in $\Sigma$. 
Consequently, there is no way to generate a considerable amount of
entropy during a bulk viscosity driven de Sitter phase. 

To investigate the stability of the solutions $H =
H_{0}$, we probe the latter with small perturbations, i.e.,   
we assume $H = H_{0}\left(1 + h\left(t\right)\right)$
with $\mid h\mid \ll 1$ in eq.(\ref{39}). 
Equation (\ref{39}) is valid for spatially flat
homogeneous and isotropic spacetimes, therefore 
the stability analysis is
restricted to this case as well. We assume that the dimensionless
quantities $f/\rho$, $\gamma$, $a$, $b$ and $v_{s}$ remain unchanged
for small deviations from $H = H_{0}$. 
Since we found the relaxation time $\tau_{0}$ in the inflationary phase
to be proportional to $H_{0}^{-1}$ it is natural to assume $\tau \sim
H^{-1}$ generally and to fix the proportionality factor by (\ref{44}). 
Retaining only terms linear in $h$, 
the resulting equation for $h$ is 
\begin{equation}
\ddot{h} + 3H_{0} K\dot{h} = 0 \K\label{55}
\end{equation}
with 
\begin{equation}
K \equiv 1 + \frac{f}{\gamma\rho} + v_{s}^{2} 
- \frac{\gamma}{2}\left(\frac{\rho}{f}f' + a\right)\p\label{56}
\end{equation}
There exists a solution $h = const$ that may be used to redefine
$H_{0}$. The other solution is stable for $K > 0$. 
In the radiation dominated case  $\gamma = 4/3$, $v_{s}^{2} = 1/3$,
$a = 1$, $b = 1/3$ and with $f = \alpha\rho$ the solution is stable for
any $\alpha > 0$. 
In the matter dominated case  $\gamma = 1$, $v_{s}^{2} \ll 1$,
$a \gg 1$ and $b = 2/3$ we find $K \approx - a/2 = - m/3T$, i.e., the
solution is unstable. 

We conclude  that there exist
stable viscosity driven inflationary solutions as long as the equation
of state is close to that for relativistic particles. 
There do not exist stable solutions for equations  of state close to
that for dust. 
Even for the stable solutions, however, 
the temperature increases exponentially
during the de Sitter stage. 
Since the particle number density decreases exponentially, this 
unfamiliar
behaviour is unavoidable to guarantee $\rho = \rho_{0}$ as long as
$\partial \rho/\partial n$ and 
$\partial \rho/\partial T$ are assumed to be positive. 
A temperature dependence like this 
is not as strange as it might appear
at the first glance. 
A nonvanishing bulk viscosity has always the tendency to heat up the
Universe. Close to local equilibrium this means that the decrease of $T$
due to the expansion of the Universe is less than without bulk
viscosity. 
Applying the cosmological dynamics with bulk viscosity to situations far
from equilibrium, in our case to a hypothetical inflationary phase, 
the $\pi$-term just compensates the equilibrium terms
in (\ref{7}) and (\ref{33}). 
It overcompensates them, however, in the case of the temperature  
(\ref{18}). 

In the following section we show how particle production processes
may modify this behaviour of the thermodynamic quantities during a de
Sitter phase. 
\section{Bulk viscous pressure and particle production}
\subsection{Basic dynamics}
Throughout this paper the Universe is studied within a single fluid
model. There is 
entropy production due to 
a nonvanishing bulk pressure. While both for pure
radiation and pure dust the bulk viscosities tend to zero, considerable
values of the latter are expected in mixtures of relativistic and
nonrelativistic matter. 
Consequently, the single fluid universe with bulk viscosity may be
regarded as a 
simplified description of a system of two (or more) interacting
components. 
Usually it is assumed that the interaction responsible for the existence
of a nonvanishing viscous pressure does not change the overall number
of fluid particles, i.e., it is assumed that (\ref{4}) holds. 
However, interactions with conserved particle numbers are only a special
case, particularly at high energies. 
Moreover, at times of the order of the Planck time or during the  
scalar field decay in inflationary scenarios, particle or string
production processes  are supposed to affect
the cosmological dynamics, leading to features like `deflationary  
universes' and `string-driven inflation' [16 - 20, 38]. 

In this section we generalize the previous formalism to the case that
the existence of a nonvanishing bulk pressure is accompanied by an
increase in the number of fluid particles. 
We start our investigations with 
(\ref{1}), (\ref{2}), (\ref{3}) and  (\ref{5}) as well, 
but (\ref{4}) is now replaced by 
\begin{equation}
N^{a}_{;a}  =  n\Gamma\K\label{57}
\end{equation}
yielding 
\begin{equation}
\dot{n} + \Theta n =  n\Gamma\p\label{58}
\end{equation}
$\Gamma$ is the particle production rate which has to be regarded as
an input quantity in our phenomenological description. 
Instead of (\ref{9}) now 
\begin{equation}
nT\dot{s} =  - \Theta\pi -\left(\rho + p\right)\Gamma
\label{59}
\end{equation}
results. 
The entropy production density is given by 
\begin{equation}
TS^{a}_{;a}  =  - n\mu \Gamma  
- \pi\left[\Theta + \frac{\tau}{\zeta}
\dot{\pi} + \frac{1}{2}\pi T\left(\frac{\tau}{\zeta T}u^{a}\right)_{;a}
\right]
\K\label{60}
\end{equation}
where $\mu$ is the chemical potential 
\begin{equation}
\mu  =  Ts - \frac{\rho + p}{n}
\p\label{61}
\end{equation}
A change in the number of particles, i.e., $\Gamma \neq 0$, is
believed to be phenomenologically equivalent to an effective viscous
pressure [13, 15, 17 - 21]. 
A discussion of this equivalence on the level of relativistic
kinetic theory has been given recently [33]. 
In order to relate the present investigations to previous work we
shall in a first step focus on the Eckart theory, i.e., to the case
$\tau = 0$. 
Equation (\ref{60}) reduces to 
\begin{equation}
TS^{a}_{E;a}  =  - n\mu\Gamma - \pi_{E}\Theta 
\label{62}
\end{equation}
in this case, 
where the subscript $E$  stands for `Eckart'. 
Formal rewriting yields 
\begin{equation}
TS^{a}_{E;a}  =  - \pi_{E}\left[\Theta + n\mu\pi_{E}^{-1}\Gamma\right]
\p\label{63}
\end{equation}
As usual, one has to guarantee $S^{a}_{E;a} \geq 0$. 
Since we expect the particle production to be effectively equivalent
to an viscous pressure we have to demand additionally that the entire
entropy production is given in terms of $\pi_{E}$, i.e., 
\begin{equation}
S^{a}_{E;a}  =  \frac{\pi_{E}^{2}}{\zeta T}  
\K\label{64}
\end{equation}
where $\pi_{E}$ is determined by 
\begin{equation}
\pi_{E}  =  - \zeta\left[\Theta + n\mu\pi_{E}^{-1}\Gamma\right]
\K\label{65}
\end{equation}
or, 
\begin{equation}
\pi_{E}^{2} + \zeta\pi_{E}\Theta  = - \zeta n\mu\Gamma 
\K\label{66}
\end{equation}
where $\zeta$ now is a generalized bulk viscosity coefficient. 
For $\Gamma \neq 0$ the latter inhomogeneous quadratic equation
replaces the familiar linear relation $\pi_{E} = - \zeta\Theta$. 
Via (\ref{66}) the particle production rate $\Gamma$ influences the
viscous pressure $\pi$. For $\Gamma = 0$ as well as for 
$\mu = 0$ we recover $\pi_{E} = - \zeta \Theta$. 
It is convenient to split $\pi_{E}$ in the following way. 
Let $\lambda$ be a not necessarily constant parameter lying in the
range 
$0 \leq \lambda \leq 1$ such that the fraction $\lambda \pi_{E}$ 
of $\pi_{E}$
describes a `creation' pressure, due to $\Gamma \neq 0$, 
while the fraction 
$\left(1 - \lambda\right)\pi_{E}$ is connected with a conventional bulk
viscosity. 
With this splitting (\ref{59}) may be written as 
\begin{equation}
nT\dot{s}_{E} = - \left(1 - \lambda\right)\Theta \pi_{E} 
- \lambda\Theta \pi_{E} - \left(\rho + p\right)\Gamma  
\p\label{67}
\end{equation}
Frequently [22 - 25] the assumption was made that the creation
process does not affect the entropy per particle. 
This is equivalent to the requirement that the terms in (\ref{67}) 
due to the creation process cancel among themselves:
\begin{equation}
\left(\rho + p\right)\Gamma = - \lambda \pi_{E}\Theta \p\label{68}
\end{equation}
Physically this means that the particles are created with a fixed 
given 
entropy. Then $\dot{s}_{E}$ is given by
\begin{equation}
nT\dot{s}_{E} = - \left(1 - \lambda\right)\Theta \pi_{E} \p\label{69}
\end{equation}
Only that fraction of $\pi_{E}$ that is not connected with a change in
the particle number contributes to $\dot{s}_{E}$. 
In the limit $\lambda = 1$, in which the entire viscous pressure is
due to particle production, one has $\dot{s}_{E} = 0$ and 
$S^{a}_{E;a} = ns_{E}\Gamma$, i.e., there is entropy production  
because of
the enlargement of the phase space. 
Using (\ref{68}) in (\ref{65}) yields 
\begin{equation}
\pi_{E}  =  - \zeta\Theta\left[1 - \lambda 
\frac{n\mu}{\rho + p}\right]
\p\label{70}
\end{equation}
Combining the latter relation again with (\ref{68}) we find that  
the part 
$\lambda\zeta$ of the generalized coefficient of bulk viscosity
$\zeta$ that arises due to a nonvanishing particle production rate is
\begin{equation}
\lambda\zeta  = \frac{\Gamma \Theta^{-2}}
{\rho + p  - \lambda n \mu} 
\p\label{71}
\end{equation}
Of course, (\ref{70}) with (\ref{64}) is also obtained if the 
condition (\ref{68}) is
immediately introduced in (\ref{62}). 

Obviously, (\ref{64}) with  (\ref{65}) is not the only way to
guarantee $S^{a}_{;a} \geq 0$. For a different approach that treats
$\Gamma$ as a thermodynamic flux independent of $\pi$, the reader is
referred to a paper by Gariel and Le Denmat [34]. 

Coming back now to the full second order theory again, we try to find
a causal evolution equation for $\pi$ that yields (\ref{12}) in the  
case 
$\Gamma \neq 0$ as well. 
Equation (\ref{60})  may be written as 
\begin{equation}
TS^{a}_{;a}  =  - \pi\left[\Theta + \frac{\tau}{\zeta}
\dot{\pi} + \frac{1}{2}\pi T\left(\frac{\tau}{\zeta
T}u^{a}\right)_{;a} + \mu\frac{n\Gamma}{\pi}\right]
\K\label{72}
\end{equation}
generalizing the relation (\ref{63}) of the Eckart theory. 
The condition $S^{a}_{;a} \geq 0$ with (\ref{12}) is now fulfilled for 
\begin{equation}
\pi^{2} + \tau \pi\dot{\pi} 
+ \frac{1}{2}\pi^{2} T\left(\frac{\tau}{\zeta
T}u^{a}\right)_{;a} + \zeta \pi \Theta = 
- \zeta \mu n\Gamma 
\K\label{73}
\end{equation}
instead of (\ref{66}).  The viscous pressure 
$\pi$ is determined by a nonlinear inhomogeneous differential
equation. For $\mu = 0$ eq.(\ref{73}) formally reduces to (\ref{11})  
For the sake of generality we shall however retain the $\mu$-term
in the following considerations. 
A chemical potential may act, e.g., as an effective symmetry breaking
parameter in relativistic field theories [35-37]. 

In the Eckart theory the general nonlinear relation (\ref{66}) was  
reduced
to the linear relation (\ref{70}) by the requirement (\ref{68}). 
Unfortunately, a corresponding simplification of (\ref{73}) via a  
relation
(\ref{68}) is {\it not} possible in general. 
The essential physical difference between the noncausal and the  
causal theory
is the appearance of a finite relaxation time within the latter. 
If a nonvanishing $\Gamma$ is responsible for the occurence of an
effective viscous pressure, a causal theory has to include a finite
relaxation time which is just the time interval during which the
corresponding part of $\pi$ decays to zero after $\Gamma$ has been
switched off. 
But the relation (\ref{68})  is an Eckart type relation due to which 
$\lambda \pi = 0$ is implied by $\Gamma = 0$ immediately. 
Therefore an approximation (\ref{68}) cannot be used in a causal  
theory. 
If however, the relation (\ref{68}) which follows from the  
requirement that the
creation process does not affect the entropy per particle has to be
abandoned, this means that within the causal theory the creation
process contributes to $\dot{s}$ in general. 
While in the Eckart theory the particle production rate was rather
simply related to the viscous pressure by  (\ref{68}),
$\Gamma$ enters the equation for $\pi$ 
of the causal theory as an independent parameter in a less obvious
way. 

Our present approach assumes that the deviations from equilibrium may
be characterized in terms of one single quantity $\pi$ also for the
case of a nonvanishing particle production rate $\Gamma$.  
This implies that there is only one (generalized) coefficient of bulk
viscosity $\zeta$ and one relaxation time $\tau$. 
The advantage of this assumption is that there exists only one causal
evolution equation, namely (\ref{73}), taking into account the
influence of $\Gamma$ on $\pi$. (For an alternative proposal see 
again [31]).  
Since the latter equation is nonlinear there is no obvious separation
into a conventional bulk pressure and a `creation' pressure. 
We shall assume furtheron, that $\zeta$ and $\tau$ continue to be
related by (\ref{28}). 
The dependence of the relaxation time 
$\tau$ on $\Gamma$ will be discussed below for
specific cases. 

For $\mu = 0$ eq.(\ref{73}) appears to be identical to the case
$\Gamma = 0$, but there are differences in the behaviour of the particle
number density and the temperature. 
The former is determined by (\ref{58}). 
Using (\ref{16}), (\ref{17}), (\ref{7}) and (\ref{58}) 
instead of (\ref{6}), 
the temperature law (\ref{18}) is replaced by (cf. [22]) 
\begin{equation}
\frac{\dot{T}}{T}  = - \Theta
\left[\frac{\partial p/\partial T}{\partial \rho/\partial T} 
+  \frac{\pi}{T \partial \rho/\partial T}\right] 
+ \Gamma 
\left[\frac{\partial p/\partial T}{\partial \rho/\partial T} 
-  \frac{\rho + p}{T \partial \rho/\partial T}\right] 
  \label{74}
\end{equation}
in the case of a nonvanishing particle production rate. 
The particle production affects the temperature not only through the
effective viscous pressure $\pi$ but there is an additional direct
coupling as well. 
Alternatively, (\ref{74}) may be written as
\begin{eqnarray}
\frac{\dot{T}}{T}  &=& - \left(\Theta - \Gamma\right)
\frac{\partial p/\partial T}{\partial \rho/\partial T} 
-  \frac{\pi\Theta + \left(\rho + p\right)\Gamma}
{T \partial \rho/\partial T} \nonumber\\
&=& 
- \left(\Theta - \Gamma\right)
\frac{\partial p/\partial T}{\partial \rho/\partial T} 
+  \frac{n\dot{s}}{\partial \rho/\partial T} 
\p  \label{75}
\end{eqnarray}
With this evolution law for the temperature, different from (\ref{18}) 
the evolution equation (\ref{73}) for 
$\pi$ becomes different from (\ref{31}) 
even in the case $\mu = 0$, although (\ref{10}) 
and (\ref{72}) seem to coincide. 
The relation (\ref{36}) for $\dot{p}$ is now replaced by 
\begin{equation}
\dot{p} = \rho \left[\left(\Gamma - \Theta\right)\gamma v_{s}^{2} 
- b\left(\gamma\Gamma + \Theta \frac{\pi}{\rho}\right)\right]
\p \label{76} 
\end{equation}
Instead of  (\ref{38}) we have
\begin{equation}
\kappa\dot{\pi} = - 2 \ddot{H} - 6 H\dot{H}\left(1 + b\right) 
+ 9H^{3}\gamma 
\left(v_{s}^{2} - b\right) - 3H^{2}\Gamma \gamma 
\left(v_{s}^{2} - b\right)
\p \label{77} 
\end{equation}
The evolution equation for $\pi$ may be written
\begin{eqnarray}
\pi + \tau\dot{\pi}&=& - \rho\tau
\left\{\frac{f}{\rho}\left(\Theta + \frac{n\mu}{\rho + p}
\frac{\gamma}{\pi/\rho}\Gamma\right)\right. \nonumber\\
&&\mbox{\ \ \ \ \ \ }+ \frac{\pi}{2\rho}\left[\Theta\left(1 + b +  
\gamma\frac{\rho f'}{f}\right) 
- \Gamma\left(b - \gamma a\right)\right] \nonumber\\
&&\left.\mbox{\ \ \ \ \ \ }+ \frac{\pi^{2}}{2\rho^{2}} \Theta
\left(a + \frac{\rho f'}{f}\right)\right\}
\K\label{78}
\end{eqnarray}
generalizing (\ref{31}).  
A procedure analogous to that leading to (\ref{39}) in section 2
provides us with an evolution equation for $H$ in a flat, homogeneous
and isotropic universe with particle production:
\begin{eqnarray}
\left\{\tau \ddot{H} - \frac{\dot{H}^{2}}{H}\tau\left(a +
\frac{\rho}{f}f'\right) 
- 3\dot{H}H\tau\left[\gamma a 
+ \frac{\gamma}{2} \frac{\rho}{f}f' 
- \frac{3}{2}\left(1 + b\right)\right]\right.&&
\nonumber\\
+ \dot{H} + \frac{1}{2}\Gamma\dot{H}\tau\left(\gamma a - b\right) 
&&\nonumber\\
- \frac{9}{2}H^{3}\tau\left[\frac{f}{\rho} 
+ \frac{\gamma}{2}\left(\gamma a - 1 - b\right)  
+ \gamma\left(v_{s}^{2} - b\right)\right] &&\nonumber\\
\left.+ \frac{3}{2}\tau H^{2}\Gamma\gamma
\left(v_{s}^{2} - b + \frac{\gamma a - b}{2}\right) 
+ \frac{3 \gamma H^{2}}{2}\right\}
\left\{- \gamma - \frac{2}{3}\frac{\dot{H}}{H^{2}}\right\}&&\nonumber\\
- \frac{3}{2}H^{2}\Gamma\tau\frac{f}{\rho}\gamma
\frac{n\mu}{\rho + p}
= 0\p&& \label{79}
\end{eqnarray}
For an easier comparison with eq.(\ref{39}) we kept separate the
common factor $\pi/\rho$ (see (\ref{35})) on the l.h.s. of this
equation. 
Having only one evolution equation for $H$ is a consequence of our
previous assumption that it is possible to characterize the
deviations from equilibrium by a single quantity $\pi$ only. 

Eq.(\ref{79}) simplifies considerably for $\mu = 0$:
\begin{eqnarray}
\tau \ddot{H} - \frac{\dot{H}^{2}}{H}\tau\left(a +
\frac{\rho}{f}f'\right) 
- 3\dot{H}H\tau\left[\gamma a 
+ \frac{\gamma}{2} \frac{\rho}{f}f' 
- \frac{3}{2}\left(1 + b\right)\right]&&
\nonumber\\
+ \dot{H} + \frac{1}{2}\Gamma\dot{H}\tau\left(\gamma a - b\right) 
&&\nonumber\\
- \frac{9}{2}H^{3}\tau\left[\frac{f}{\rho} 
+ \frac{\gamma}{2}\left(\gamma a - 1 - b\right)  
+ \gamma\left(v_{s}^{2} - b\right)\right] &&\nonumber\\
+ \frac{3}{2}\tau H^{2}\Gamma\gamma
\left(v_{s}^{2} - b + \frac{\gamma a - b}{2}\right) 
+ \frac{3}{2}\gamma H^{2}
\stackrel{\left(\mu = 0\right)}{=} 0\p&& \label{80}
\end{eqnarray}
For $\Gamma = 0$ both equations reduce to (\ref{39}).   
Eliminating $\pi$ in the evolution equation (\ref{74})
for $T$ with the help of
(\ref{35}), we find 
\begin{equation}
\frac{\dot{T}}{T} = 3\left(\gamma a - b\right)\frac{\dot{R}}{R} 
- \left(\gamma a - b\right)\Gamma  
+ a \frac{\dot{\rho}}{\rho} \p\label{81}
\end{equation}
instead of (\ref{41}).   
For constant values of $a$, $b$ and $\gamma$ 
\begin{equation}
T \sim \rho^{a}\left(\frac{R^{3}}{N}\right)^{\gamma a - b}   
= \rho^{a}\left(\frac{1}{n}\right)^{\gamma a - b}
\p\label{82}
\end{equation}
For $\gamma = 4/3$, $b = 1/3$, $a = 1$, this specifies to 
\begin{equation}
T \sim \rho\frac{R^{3}}{N} = \frac{\rho}{n}
\p\label{83}
\end{equation}
\subsection{Inflationary solutions}
Looking for solutions $H = H_{0} = const$ of (\ref{79}),  
we find the following expression for $\tau_{0}$ that generalizes
(\ref{44}) 
\begin{eqnarray}
\tau_{0}^{-1} &=& 3H_{0}
\left[\frac{f}{\gamma\rho}  
+ \frac{\gamma a - 1 - b}{2}
+ v_{s}^{2} - b\right] \nonumber\\
&& - \Gamma \left[\frac{n\mu}{\rho + p}\frac{f}{\gamma\rho}  
+ \frac{\gamma a - b}{2}
+ v_{s}^{2} - b\right] 
\p\label{84}
\end{eqnarray}
If the particle production rate is comparable with the expansion
rate, it may essentially influence the relaxation time $\tau_{0}$. 
Especially interesting is the possibility that the resulting effect
of the $\Gamma$-terms in (\ref{84}) is to enlarge $\tau_{0}$. 
In this case particle production may either enable or improve the
fulfilment of the `freezing in' condition $\tau_{0} > H_{0}^{-1}$ for
the nonequilibrium. 
In other words, the consistency criterion for inflation 
might be easier to fulfill in the case with particle production than
without. 
Of course, these considerations make only sense as long as $\tau_{0}$
remains finite. As we shall see below there are parameter
combinations for which $\tau_{0}$ diverges. 

If during some time interval the particle production rate $\Gamma$ is
proportional to the expansion rate and approximately constant as
well, i.e., $\Gamma = 3\lambda H_{0}$, we get 
\begin{equation}
\tau_{0}^{-1} = 3H_{0}
\left[\frac{f}{\gamma\rho}\left(1 - \frac{\lambda n\mu}{\rho +
p}\right) - \frac{1}{2} + \left(1 - \lambda\right)
\left(\frac{\gamma a - b}{2}
+ v_{s}^{2} - b\right)\right]
\p\label{85}
\end{equation}
For particles with $\mu = 0$ and an equation of state close to that
for radiation it is obvious that for any $\lambda > 0$ the relaxation
time $\tau_{0}$ is larger than for $\lambda = 0$. 
In the limiting case $\lambda = 1$ the production rate of the
particles coincides with the expansion rate. $\tau_{0}$ then is given
by 
\begin{equation}
\tau_{0}^{-1} \stackrel{\left(\lambda = 1\right)}{=} 3H_{0}
\left[\frac{f}{\gamma\rho}\left(1 - \frac{n\mu}{\rho +
p}\right) - \frac{1}{2}\right]
\p\label{86}
\end{equation}
Specifying again to $\gamma = 4/3$ and restricting ourselves to $\mu
= 0$,  
\begin{equation}
\tau_{0}^{r} \stackrel{\left(\lambda = 1\right)}{=} 
\frac{4}{9}\frac{H_{0}^{-1}}{\frac{f}{\rho} - \frac{1}{2}}
\label{87}
\end{equation}
results. 
For $f = \rho$, the usual choice in the literature [10, 28, 29, 31],
this expression for $\tau_{0}$ is twice as large as the corresponding
expression from (\ref{45}), although $\tau_{0}^{r} < H_{0}^{-1}$ in 
both cases. 
For $f = 2\rho/3$ however, we find $\tau_{0}^{r} < H_{0}^{-1}$ 
from (\ref{45}), while (\ref{87}) yields 
$\tau_{0}^{r} > H_{0}^{-1}$. 
In the latter case the nonequilibrium is `frozen in', in the former
one it is not. 
This demonstrates explicitly that under certain circumstances 
particle production may improve the conditions for bulk viscous
inflation. 
On the other hand,  (\ref{87}) makes sense only for $f/\rho > 1/2$  
since for
$f = \rho/2$ the relaxation time diverges. 
Especially the case $f = \rho/3$, dealt with for $\Gamma = 0$ in
section 2, is impossible for $\Gamma = 3H_{0}$. 

The case $\Gamma = 3H_{0}$ is singled out in different respects as
well. 
For any $\Gamma < 3H_{0}$ the particle number density $n$ is
decreasing according to (\ref{58}), while according to (\ref{82}) the
temperature $T$ increases correspondingly to guarantee 
$\dot{\rho}_{0} = \dot{H}_{0} = 0$. 
In the limit $\Gamma = 3 H_{0}$, $n$ becomes constant 
and it follows from (\ref{82}) that the temperature is constant as  
well. 
Only in this extreme limiting case which corresponds to $\dot{s} = 0$
(this is only possible since we restricted ourselves to a time
interval with $\Gamma = const$), the temperature may remain constant
in a de Sitter phase driven by an effective bulk pressure. 
For  $\Gamma < 3H_{0}$ with $\dot{s} \neq 0$ the temperature
neccessarily increases, although for $\Gamma \neq 0$ at a lower rate
than for $\Gamma = 0$. 
\subsection{Entropy production}
Eliminating $\pi$ from (\ref{59}) with the help of (\ref{35}) 
and  (\ref{40}) the time
dependence of the entropy per particle is determined by
\begin{equation}
nT\dot{s} = \left(\rho + p\right)
\left[3H - \Gamma\right] + \dot{\rho}
\K\label{88}
\end{equation}
or
\begin{equation}
nT\dot{s} = \left(\frac{\rho R^{3\gamma}}{N^{\gamma}}
\right)^{\displaystyle \cdot} \frac{N^{\gamma}}{R^{3\gamma}}
\p\label{89}
\end{equation}
With (\ref{58}) and 
\begin{equation}
T= T_{0}\left(\frac{\rho}{\rho_{0}}\right)^{a}
\left(\frac{R^{3} N_{0}}{R_{0}^{3} N}\right)^{\gamma a - b} 
\label{91}
\end{equation}
from (\ref{82}), where the subscript $0$ refers to some initial time, 
one finds 
\begin{equation}
\dot{s} = \frac{1}{n_{0}T_{0}}
\left(\frac{\rho_{0}}{\rho}\right)^{a}
\left(\frac{R_{0}^{3} N}{R^{3} N_{0}}\right)^{\gamma a - 1 - b} 
\left(\frac{N}{R^{3}}\right)^{\gamma}
\left(\frac{\rho R^{3\gamma}}{N^{\gamma}}
\right)^{\displaystyle \cdot} \p\label{92}
\end{equation}
In the inflationary phase with $\rho = \rho_{0}$ and $H = H_{0}$, 
we have 
\begin{equation}
\dot{s} = \frac{\rho_{0}}{n_{0}T_{0}}\gamma
\left(\frac{R_{0}^{3} N}{R^{3} N_{0}}\right)^{\gamma a - 1 - b} 
\left[3 H_{0} - \Gamma\right]
\K\label{93}
\end{equation}
which for radiation ($\gamma = 4/3$, $a = 1$, $b = 1/3$) reduces to 
\begin{equation}
\dot{s} = \frac{4\rho_{0}}{n_{0}T_{0}}
\left[H_{0} - \frac{1}{3}\Gamma\right]
\p\label{94}
\end{equation}
Integration of (\ref{94}) yields 
\begin{equation}
s = \frac{4\rho_{0}}{n_{0}T_{0}}
\left[H_{0}\left(t - t_{0}\right) 
- \frac{1}{3}\int_{0}^{t}\Gamma dt\right] + s\left(t_{0}\right) 
\p\label{95}
\end{equation}
The basic difference compared with the case $\Gamma = 0$ is that the
change of the entropy in a comoving volume, $\Sigma = nsR^{3}$, 
is no longer determined by the change of $s$ alone. 
With the equations of state for radiation, $\Sigma$ is given by 
\begin{equation}
\Sigma = \left\{\frac{4\rho_{0}}{n_{0}T_{0}}
\left[H_{0}\left(t - t_{0}\right) 
- \frac{1}{3}\int_{0}^{t}\Gamma dt\right] + s\left(t_{0}\right)\right\}
N_{0}\exp{\int_{0}^{t}\Gamma dt}
\label{96}
\end{equation}
in the inflationary phase. 
If $\Gamma$ is again assumed to be approximately given by 
$\Gamma = 3\lambda H_{0}$, the expression (\ref{96}) reduces to 
\begin{equation}
\Sigma\left(t\right)  = 
\left[\frac{4\rho_{0}}{n_{0}T_{0}}\left(1 - \lambda\right)
H_{0}\left(t -
t_{0}\right) + s\left(t_{0}\right)\right]
N\left(t_{0}\right)
\exp{\left[3\lambda H_{0}\left(t - t_{0}\right)\right]}
\p \label{97}
\end{equation}
It follows that for $\Gamma \neq 0$ we have 
an exponential increase of the comoving entropy $\Sigma$ during the
de Sitter phase. 
Starting from the latter expression for $\Sigma$ 
it is possible to apply Maartens' numerical
estimations [29] concerning the entropy production during the  
inflationary
phase. All his considerations of this point are valid in the present 
case for $\lambda \neq 0$, provided his $H_{0}$ is replaced by
$\lambda H_{0}$. 
His conclusion that it is possible `to generate the right amount of
entropy without re-heating' may hold in a universe with particle
production. 
There is complete equivalence to Maartens' result for the entropy
production 
for $\lambda = 1$. 
Consequently, a substantial production of entropy during a (bulk
viscosity induced) dissipational inflationary phase is possible if at
least a part of the bulk viscous pressure is related to an increase
in the particle number. 
In a sense we have reestablished Maartens' result, although within a
different setting: 
An exponential
growth of the comoving entropy is only possible if the viscous
pressure $\pi$ in (\ref{1}) is, 
at least partially, a `creation' pressure connected with an
increase in the number of fluid particles rather than an increase in
the entropy per particle. 

It remains to consider the stability of the inflationary
solutions for $\Gamma \neq 0$. Proceeding as in section 2 we have 
\begin{equation}
\ddot{h} + \left(3 H_{0} K - \Gamma M\right)\dot{h}  = 0
\label{98}
\end{equation}
instead of (\ref{55}), where 
\begin{equation}
M \equiv 2 \frac{f}{\gamma\rho}\frac{n\mu}{\rho + p}  
+ v_{s}^{2} - b 
\p\label{99}
\end{equation}
Consequently, the inflationary solutions are stable for 
$3 H_{0} K - \Gamma M > 0$. 
For $\mu = 0$ and the equations of state for radiation one has $M =
0$ and the stability condition reduces to $K > 0$ again. 
The stability properties of the corresponding inflationary solutions
are not affected at all by a nonvanishing $\Gamma$. 
With equations of state for matter the inflationary solutions have
the same instability as for $\Gamma = 0$. 
\section{Summary}
Using particle number density and temperature as basic
thermodynamical variables of the cosmic fluid 
we have studied the full M\"uller-Israel-Stewart theory for a
spatially flat, homogeneous and isotropic universe with 
bulk viscous pressure. 
We found general criteria for the applicability of the so called
truncated versions. 
It was shown that in exceptional cases there exist common solutions
of the full and the truncated theories far from equilibrium. 
The possibility of exponential inflationary solutions driven by 
bulk viscosity  was investigated.  Almost 
all corresponding solutions imply an exponential growth of the
temperature during the de Sitter stage. 
Since the number density of the fluid particles decreases
exponentially, only a corresponding increase in the temperature
guarantees a constant energy density as long as the specific heat is
positive and finite and the energy density  increases 
with the particle number density. 
In the second part of the paper the bulk viscous pressure was allowed
to account partially or fully for particle production
processes. 
A nonvanishing particle production rate may enlarge the relaxation
time for a viscous pressure that is supposed to drive inflation. 
It may help to `freeze in' the corresponding nonequilibrium, i.e., to
improve the conditions for inflation. 
There exist stable, inflationary solutions for equations of state
close to that for radiation. 
Due to the increase in the particle number 
the comoving entropy  increases exponentially in this  period. 
Only in the limiting case that the entire inflation driving 
viscous pressure is
due to particle production 
both the number density and the temperature
remain constant during the de Sitter phase.

\ \\
{\bf Acknowledgment}\\
This paper was inspired by a preprint of [29]. I thank Roy Maartens, 
Portsmouth, for clarifying conversations. 
I am indebted to the members of the Grup de F\'{\i}sica
Estad\'{\i}stica of the Autonomous University of Barcelona 
for their warm hospitality. Helpful discussions with
Diego Pav\'on who critically commented on previous versions of this
paper, David Jou and Josep Triginer are gratefully acknowledged.

\ \\
{\bf References}\\
\ \\
\ [1] C. Eckart, {\it Phys. Rev.} {\bf 58}, 919 (1940).\\
\ [2] L.D. Landau and E.M. Lifshitz,  
{\it Fluid Mechanics}, \\
$\mbox{\ \ \ \ \ }$(Addison-Wesley, Reading, 1958).\\
\ [3] I. M\"uller, {\it Z. Physik} {\bf 198}, 329 (1967).\\
\ [4] W. Israel, {\it Ann. Phys. (NY)} {\bf 100}, 310 (1976).\\
\ [5] W. Israel and J.M. Stewart, {\it Ann. Phys. (NY)} {\bf 118},  
341 (1979).\\
\ [6] W. Israel and J.M. Stewart,  
{\it Proc. R. Soc. Lond. A} {\bf 365}, 43 (1979).\\
\ [7] D. Pav\'{o}n, D. Jou  and J. Casas-V\'{a}zquez,  \\
$\mbox{\ \ \ \ \ }$
{\it Ann. Inst. H. Poincar\'{e}} Sec. A {\bf 36}, 79 (1982).\\
\ [8] W.A. Hiscock  and L. Lindblom, {\it Ann. Phys. (NY)} {\bf 151},
466 (1983).\\
\ [9] W.A. Hiscock  and L. Lindblom,  
{\it Phys. Rev. D} {\bf 35}, 3723 (1987).\\
\newpage
\ \\
\ [10] V.A. Belinskii, E.S. Nikomarov and I.M. Khalatnikov,  \\
$\mbox{\ \ \ \ \ }$
{\it Sov. Phys. JETP} {\bf 50}, 213 (1979).\\
\ [11] {\O}. Gr{\o}n, 
{\it Astrophys. Space Sci.} {\bf 173}, 191 (1990).\\
\ [12] N. Udey and W. Israel,  
{\it Mon. Not. R. Astr. Soc.} {\bf 199}, 1137 (1982).\\
\ [13] Ya.B. Zel'dovich,   
{\it Sov. Phys. JETP Lett.} {\bf 12}, 307 (1970).\\
\ [14] G.L. Murphy,
{\it Phys. Rev. D} {\bf 8}, 4231 (1973).\\
\ [15] B.L. Hu, 
{\it Phys. Lett. A} {\bf 90}, 375 (1982).\\
\ [16] N. Turok, 
{\it Phys. Rev. Lett} {\bf 60}, 549 (1988).\\
\ [17] J.D. Barrow,  
{\it Phys. Lett. B} {\bf 180}, 335 (1986).\\
\ [18] J.D. Barrow,  
{\it Phys. Lett. B} {\bf 183}, 285 (1987).\\
\ [19] J.D. Barrow,  
{\it Nucl. Phys. B} {\bf 310}, 743 (1988).\\
\ [20] J.D. Barrow in {\it The Formation and Evolution of Cosmic  
Strings}\\
$\mbox{\ \ \ \ \ }$
edited by G. Gibbons, S.W. Hawking and T. Vachaspati\\
$\mbox{\ \ \ \ \ }$
(Cambridge University Press, Cambridge, 1990), p. 449.\\
\ [21] I. Prigogine, J. Geheniau, E. Gunzig and P. Nardone  \\
$\mbox{\ \ \ \ \ }$ 
{\it Gen. Rel. Grav.} {\bf 21}, 767 (1989).\\
\ [22] M.O. Calv\~{a}o, J.A.S. Lima and I. Waga, 
{\it Phys. Lett. A} {\bf 162}, 223 (1992).\\
\ [23] W. Zimdahl and D. Pav\'on,
{\it Phys. Lett. A} {\bf 176}, 57 (1993).\\
\ [24] W. Zimdahl and D. Pav\'on, 
{\it Mon. Not. R. Astr. Soc.} {\bf 266}, 872 (1994).\\
\ [25] W. Zimdahl and D. Pav\'on, 
{\it Gen. Rel. Grav.} {\bf 26}, 1259 (1994).\\
\ [26] W.A. Hiscock and J. Salmonson, 
{\it Phys. Rev. D} {\bf 43}, 3249 (1991).\\
\ [27] D. Pav\'on, J. Bafaluy and D. Jou, 
{\it Class. Quantum Grav.} {\bf 8}, 347 (1991).\\
\ [28] M. Zakari and D. Jou, 
{\it Phys. Rev. D} {\bf 48}, 1597 (1993).\\
\ [29] R. Maartens, {\it Class. Quantum Grav.} {\bf 12}, 1455 (1995).\\
\ [30] J. Gariel and G. Le Denmat G, 
{\it Phys. Rev. D} {\bf 50}, 2560 (1994).\\
\ [31] V. Romano  and D. Pav\'on, 
{\it Phys. Rev. D} {\bf 50}, 2572 (1994).\\
\ [32] D.V. Nanopoulos, {\it Prog. Part. Nucl. Phys.} {\bf 6}, 23  
(1981),\\
$\mbox{\ \ \ \ \ }$
G. B\"orner, 
{\it The Early Universe}, chapter 8.3,\\
$\mbox{\ \ \ \ \ }$
(Springer Verlag, Berlin, 1993).\\
\ [33] J. Triginer, W. Zimdahl and D. Pav\'{o}n,  \\
$\mbox{\ \ \ \ \ }$
{\it Class. Quantum Grav.} (1996) in press.\\
\ [34] J. Gariel and G. Le Denmat, 
{\it Phys. Lett. A} {\bf 200}, 11 (1995).\\
\ [35] J.I. Kapusta, 
{\it Phys. Rev. D} {\bf 24}, 426 (1981).\\
\ [36] H.E. Haber and H.A. Weldon, 
{\it Phys. Rev. D} {\bf 25}, 502 (1982).\\
\ [37] J.A.S. Lima, R. Portugal and I. Waga, 
{\it Phys. Rev. D} {\bf 37}, 2755 (1988).\\
\ [38] J.A.S. Lima and J.M.F. Maia, 
{\it Phys. Rev. D} {\bf 49}, 5597 (1994).\\

\end{document}